\documentclass{article}
\usepackage{spconf,amsmath,graphicx}
\usepackage{amsmath}
\usepackage{amsfonts}
\usepackage{amssymb}
\usepackage{algorithm}
\usepackage{algorithmic}
\usepackage{graphicx}
\usepackage{verbatim}

\newtheorem{theorem}{Theorem}
\newtheorem{lemma}{Lemma}

\newtheorem{corollary}{Corollary}
\newtheorem{example}{Example}
\newtheorem{definition}{Definition}

\newtheorem{observation}{Observation}

\newcommand{\beq}{\begin{equation}}
\newcommand{\eeq}{\end{equation}}
\newcommand{\beqnn}{\begin{equation*}}
\newcommand{\eeqnn}{\end{equation*}}
\newcommand{\beqy}{\begin{eqnarray}}
\newcommand{\eeqy}{\end{eqnarray}}
\newcommand{\beqynn}{\begin{eqnarray*}}
\newcommand{\eeqynn}{\end{eqnarray*}}
\newcommand{\bit}{\begin{itemize}}
\newcommand{\eit}{\end{itemize}}
\newcommand{\ben}{\begin{enumerate}}
\newcommand{\een}{\end{enumerate}}
\newcommand{\bex}{\begin{example}}
\newcommand{\eex}{\end{example}}

\newcommand{\balg}[1]{\begin{algorithm} \caption{#1}}
\newcommand{\ealg}{\end{algorithm}}

\newcommand{\balgc}{\begin{algorithmic}[1]}
\newcommand{\ealgc}{\end{algorithmic}}

\newcommand{\bary}{\begin{array}}
\newcommand{\eary}{\end{array}}
\newcommand{\bmx}{\begin{bmatrix}}
\newcommand{\emx}{\end{bmatrix}}
\newcommand{\bsmx}{\left[\begin{smallmatrix}}
\newcommand{\esmx}{\end{smallmatrix}\right]}
\newcommand{\bmxc}[1]{\left[\begin{array}{@{}#1@{}}}
\newcommand{\emxc}{\end{array}\right]}
\newcommand{\bcn}{\begin{center}}
\newcommand{\ecn}{\end{center}}

\newcommand{\A}{\boldsymbol{A}}

\newcommand{\I}{\boldsymbol{I}}

\newcommand{\rr}{\boldsymbol{r}}

\newcommand{\x}{{\boldsymbol{x}}}
\newcommand{\y}{{\boldsymbol{y}}}

\newcommand{\0}{{\boldsymbol{0}}}

\def\x{{\mathbf x}}

\DeclareMathOperator*{\argmin}{arg\,min}
\DeclareMathOperator*{\argmax}{arg\,max}
\usepackage{cite}

\title{Exact Sparse Signal Recovery  via Orthogonal Matching Pursuit with Prior Information}
\name{Author(s) Name(s)
\thanks{This work was partially supported by  National Natural Science Foundation of China (No. 11871248)
and ``the Fundamental Research Funds for the Central Universities'' (No. 21618329).}}
\address{Author Affiliation(s)}
\twoauthors
  {Jinming Wen \sthanks{This work was partially supported by  NSFC (No. 11871248)
  and ``the Fundamental Research Funds for the Central Universities'' (No. 21618329).}}
	{College of Cyber Security\\
	Jinan University, Guangzhou, China\\
	E-mail: jinming.wen@mail.mcgill.ca}
  {Wei Yu 
  }
	{Dept. of Electrical
and Computer Engineering\\
	University of Toronto, Toronto,  Canada\\
	E-mail: weiyu@comm.utoronto.ca}
\begin{document}
%
\maketitle
\begin{abstract}
The orthogonal matching pursuit (OMP) algorithm is a commonly used algorithm
for recovering $K$-sparse signals $\x\in \mathbb{R}^{n}$ from linear model $\y=\A\x$,
where $\A\in \mathbb{R}^{m\times n}$ is a sensing matrix.
A fundamental question in the performance analysis of OMP is
the characterization of the probability that it can exactly recover $\x$ for random matrix $\A$.
Although in many practical applications, in addition to the sparsity,
$\x$ usually also has some additional property (for example, the nonzero entries
of $\x$ independently and identically follow the Gaussian distribution),
none of existing analysis uses these properties to answer the above question.
In this paper, we first show that the prior distribution information of $\x$ can be used to
provide an upper bound on $\|\x\|_1^2/\|\x\|_2^2$,
and then explore the bound to develop a better lower bound on the probability of exact recovery with
OMP in $K$ iterations.
Simulation tests are presented to illustrate the superiority of the new bound.
\end{abstract}
\section{Introduction}

In many applications, such as sparse activity detection \cite{CheSY18},
we need to reconstruct a $K$-sparse signal $\x$
(i.e., $\x$ has at most $K$ nonzero entries) from linear measurements:
\beq
\label{e:model}
\y=\A\x,
\eeq
where $\A\in \mathbb{R}^{m\times n}$ ($m\ll n$) is a random sensing matrix
with independent and identically distributed (i.i.d.) Gaussian  $\mathcal{N}(0,1/m)$ entries and
$\y\in \mathbb{R}^m$ is a given observation vector.
Numerous sparse recovery algorithms have been developed to recover $\x$ based on $\y$ and $\A$
\cite{CanT05,Don06, CanRT06b}. Among them, greedy algorithms are very popular,
especially when $m,n$ and/or $K$ are large, due to their low computational complexities.
The orthogonal matching pursuit (OMP) algorithm~\cite{pati1993orthogonal}, which is described in Algorithm~\ref{a:OMP},
is a widely-used greedy algorithm due to its high efficiency and effectiveness \cite{TroG07}.

A fundamental question in the analysis of OMP is the characterization of its exact recovery capability.
To this end, numerous works studied the recovery performance of OMP (see, e.g.,
\cite{Zha11,  DonH01,Tro04, DavW10, MoS12, WenZL13, WenZWM17}).
In particular, \cite{TroG07b} develops a lower bound on the probability of exact recovery
of $K$-sparse $\x$ with $K$ iterations of OMP.
To better understand its recover capability,
it is natural to ask whether this lower bound can be improved.

In many practical applications, in addition to sparsity, $\x$ also has some other properties.
For example, in wireless communication problems involving the Rayleigh channel model,
the nonzero entries of $\x$ independently and identically follow the standard Gaussian distribution $\mathcal{N}(0, 1)$
\cite{CheSY18}. In speech communication \cite{HabGC09} and  audio source separation \cite{VinBGB14},
$\x$ has exponentially decaying property, i.e., $\x$ is a $K$-sparse $\alpha$-strongly-decaying signal.
Intuitively, a larger variation in the magnitudes of the nonzero entries of $\x$ would typically
lead to better exact recovery performance of OMP in $K$ iterations.

This paper aims to develop a theoretical framework to capture the dependence of the exact recovery performance of OMP on the disparity in the magnitudes of the nonzero entries of $\mathbf{x}$. Toward this end, we define the following measure of the disparity, in term of a function $\phi(t)$, such that
\beq
\label{e:csk}
\|\x_{\mathcal{S}}\|_{1}^2\leq \phi(|\mathcal{S}|)
\|\x_{ \mathcal{S}}\|_{2}^2
\eeq
for any set $\mathcal{S}\subseteq \Omega$, where $\Omega$ is the support of $\x$,
$|\mathcal{S}|$ denotes the number of elements of $\mathcal{S}$
 and $\phi(t)$ is a nondecreasing function of $t>0$ with $0<\phi(t)\leq t$.
Note that by the Cauchy-Schwarz inequality, \eqref{e:csk} with $\phi(t)=t$ holds for any $K$-sparse
signal $\x$. Furthermore, \eqref{e:csk} with $\phi(t)$  much smaller than $t$ holds for
$\alpha$-strongly-decaying signals and random signals (more details will be provided in Sec. \ref{s:main}).

In this paper, we develop a lower bound on the probability of the exact recovery for
$K$-sparse signals $\x$ that satisfy \eqref{e:csk}, using $K$-iterations of OMP,
as a function of $\phi(t)$. Since the bound depends on the function $\phi(t)$, we develop closed-form
expressions of $\phi(t)$ for general $K$-sparse signals, $K$-sparse $\alpha$-strongly-decaying signals,
and $K$-sparse signals whose nonzero entries independently and identically follow the $\mathcal{N}(0, 1)$
distribution, leading to exact lower bounds for these three classes of sparse signals.

\begin{algorithm}[t]
\caption{The OMP Algorithm~\cite{pati1993orthogonal}}  \label{a:OMP}
Input: $\y$, $\A$, and stopping rule.\\
Initialize: $k=0, \rr^0=\y, \mathcal{S}_0=\emptyset$.\\
until the stopping rule is met
\begin{algorithmic}[1]
\STATE $k=k+1$,
\STATE $s^k=\argmax\limits_{1\leq i\leq n}|\langle \rr^{k-1},\A_i\rangle|$,
\STATE $\mathcal{S}_k=\mathcal{S}_{k-1}\bigcup\{s^k\}$,
\STATE $\hat{\x}_{\mathcal{S}_k}=\argmin\limits_{\x\in \mathbb{R}^{|\mathcal{S}_k|}}\|\y-\A_{\mathcal{S}_k}\x\|_2$,
\STATE $\rr^k=\y-\A_{\mathcal{S}_k}\hat{\x}_{\mathcal{S}_k}$.
\end{algorithmic}
Output: $\hat{\x}=\argmin\limits_{\x: \text{supp}(\x)=\mathcal{S}_k}\|\y-\A\x\|_2$.
\end{algorithm}

\section{Main Results}
\label{s:main}

In the following, we provide a lower bound on the probability
that OMP can exactly recover any $K$-sparse signal $\x$ satisfying \eqref{e:csk} in $K$ iterations
for random sensing matrix $\A$. 

\begin{theorem}
\label{t:noiseless}
Let $\A\in \mathbb{R}^{m\times n}$ be a random matrix with i.i.d. $\mathcal{N}(0, 1/m)$  entries,
and $\x$ be a $K$-sparse signal that satisfies \eqref{e:csk} for some particular $\phi(t)$.
Define the event $\mathbb{S}$ as
\beq
\label{e:E}
\mathbb{S}=\{\mbox{OMP can exactly recover $\x$ in \eqref{e:model} in $K$ iterations}\}.
\eeq
Denote interval $\mathcal{I}=\left(0,1-\sqrt{\frac{K}{m}}-\sqrt{\frac{2\phi(K)}{m\pi}}\right]$, then
\begin{align}
\label{e:prob}
\mathbb{P}(\mathbb{S})\geq\max_{\epsilon\in\mathcal{I}}(1-e^{-\frac{\epsilon^2m}{2}})
\prod_{k=1}^{K}\left(1-\frac{e^{-\frac{\eta^2m}{2\phi(k)}}}{\sqrt{\frac{\pi m}{2\phi(k)}}\eta}\right)^{(n-K)},
\end{align}
where $\eta=1-\sqrt{K/m}-\epsilon.$
\end{theorem}

The proof of Theorem \ref{t:noiseless} can be found in the journal version of this paper.
In the following, we give some remarks.

Theorem \ref{t:noiseless}  is important from both theoretical and practical applications
points of view.
Theoretically, Theorem \ref{t:noiseless} characterizes the recovery performance of OMP.
In practical applications, we can use $\eqref{e:prob}$ to give a lower bound on $\mathbb{P}(\mathbb{S})$.
If the lower bound is large, saying close to 1, then we are confident to use the OMP algorithm
to do the reconstruction.
If the lower bound is small, saying much smaller than 1, then another more effective recovery algorithm
(such as the basis pursuit \cite{CanT05}) may need to be used.

As far as we know, Theorem \ref{t:noiseless} gives the first lower bound on $\mathbb{P}(\mathbb{S})$ by using the
extra information (i.e., inequality \eqref{e:csk}) of the $K$-sparse signal $\x$.
Note that \cite[Theorem 6]{TroG07b} also gives a lower bound on $\mathbb{P}(\mathbb{S})$,
but it only uses the $K$-sparsity property of $\x$.
Since Theorem \ref{t:noiseless} uses not only the sparsity of $\x$ but also its additional property
\eqref{e:csk} to derive the lower bound, it provides a sharper lower bound on $\mathbb{P}(\mathbb{S})$ than
\cite[Theorem 6]{TroG07b}.
More details on the comparison of the two lower bounds are presented in Sec. \ref{s:sim}.

Theorem \ref{t:noiseless} can theoretically explain that OMP has better
recovery ability in recovering sparse signals with larger variation of the magnitudes of
their nonzero entries.
Specifically, it is not hard to see that the right-hand side of \eqref{e:prob} becomes larger
as $\phi(t)$ (or equivalently $\frac{\|\x_{\mathcal{S}}\|_{1}}{\|\x_{\mathcal{S}}\|_{2}}$ (see \eqref{e:csk})) becomes smaller. By the Cauchy-Schwarz inequality,
$\frac{\|\x_{\mathcal{S}}\|_{1}}{\|\x_{\mathcal{S}}\|_{2}}$
achieves the maximal value $\sqrt{|\mathcal{S}|}$
when the magnitudes of all the entries of $\x_{\mathcal{S}}$ are the same.
Hence, the probability of exact recovery of $K$-sparse $\x$, whose non-zero entries have identical magnitudes, has the smallest lower bound.
On the other hand, if the variation of the magnitudes of the nonzero entries of $\x$ are large,
then $\frac{\|\x_{\mathcal{S}}\|_{1}}{\|\x_{\mathcal{S}}\|_{2}}$
is small, and hence the right-hand side of \eqref{e:prob} is large.
Therefore, generally speaking, the probability of the exact recovery of this kind of $K$-sparse signals $\x$ is large.

As \eqref{e:prob} depends on $\phi(t)$, to lower bound $\mathbb{P}(\mathbb{S})$,
we need to know $\phi(t)$. In the following, we give closed-form expressions of $\phi(t)$
for three cases. We begin with the first case where we only know that $\x$ is $K$-sparse.
By the Cauchy-Schwarz inequality, one can see that \eqref{e:csk} holds if $\phi(t)=t$.
Hence, by Theorem \ref{t:noiseless}, we have
\begin{corollary}
\label{c:regular}
Let $\A\in \mathbb{R}^{m\times n}$ be a random matrix with i.i.d. $\mathcal{N}(0, 1/m)$ entries
and $\x$ be an arbitrary $K$-sparse signal. Then \eqref{e:prob} holds with $\phi(t)=t$.
\end{corollary}

Note that \cite[Theorem 6]{TroG07b} shows that
\begin{align}
\label{e:probtrop}
&\mathbb{P}(\mathbb{S})\nonumber\\
\geq&\max_{\epsilon\in(0,\sqrt{m/K}-1)}(1-e^{-\frac{\epsilon^2m}{2}})
\left(1-e^{-\frac{(\sqrt{m/K}-1-\epsilon)^2}{2}}\right)^{K(n-K)}
\end{align}
where the event $\mathbb{S}$ is defined in \eqref{e:E}.
Since the lower bounds on $\mathbb{P}(\mathbb{S})$ given by Corollary \ref{c:regular} and \eqref{e:probtrop}
are complicated, it is difficult to theoretically compare them.
However, from the simulation results in Sec. \ref{s:sim}, one can see that the new bound
given by Corollary \ref{c:regular} is  much sharper than that given by \eqref{e:probtrop}.

Next, we give a lower bound on $\mathbb{P}(\mathbb{S})$ for recovering $\alpha$-strongly-decaying signal.
First, we state the precise definition of $\alpha$-strongly-decaying signals as follows:
\begin{definition}
 [\cite{DavW10}]
 \label{d:alphastr}
Without loss of generality, let all the entries of $K$-sparse $\x$ be ordered as
\[
|x_1|\geq |x_2|\geq\ldots\geq |x_K|\geq 0, \, x_j=0, \mbox{ for } K+1\leq j\leq n.
\]
Then $\x$ is called as a $K$-sparse $\alpha$-strongly-decaying signal
($\alpha>1$) if
$
|x_i|\geq \alpha|x_{i+1}|, \, 1\leq i\leq K-1.
$
\end{definition}

The following lemma provides a closed-form expression of $\phi(t)$
for $K$-sparse $\alpha$-strongly-decaying signals.
\begin{lemma}
\label{l:decaying}
Let $\x$ be a $K$-sparse $\alpha$-strongly-decaying signal, then \eqref{e:csk} holds with
\begin{align}
\label{e:phit}
\phi(t)=\frac{(\alpha^t-1)(\alpha+1)}{(\alpha^t+1)(\alpha-1)}, \,\; t>0.
\end{align}
\end{lemma}

The proof of Lemma \ref{l:decaying} can be found in the journal version of this paper.

%
%

\begin{figure}[t]
\centering
\includegraphics[width=3.0in]{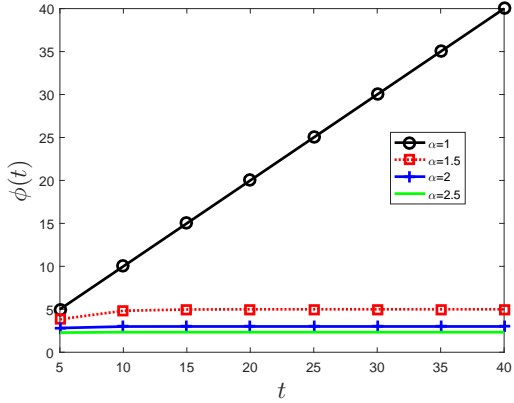}
\caption{$\phi(t)$ versus $\alpha$ with $\alpha=1, 1.5,2, 2.5$}
\label{f:phi1}
\end{figure}

To show how large the $\phi(t)$ in \eqref{e:phit} is, we plot it for different values of $\alpha$
in Fig.\ref{f:phi1}, where for comparison, we also plot $\phi(t)=t$
(note that this is equivalent to the $\alpha=1$ case as $\lim\limits_{\alpha\rightarrow1}\phi(t)=t$).
Fig. \ref{f:phi1} shows that $\phi(t)$ is much smaller than $t$ for
large $t$ and/or $\alpha$.

Theorem \ref{t:noiseless} and Lemma \ref{l:decaying} implies the following corollary:
\begin{corollary}
\label{c:decay}
Let $\A\in \mathbb{R}^{m\times n}$ be a random matrix with i.i.d. $\mathcal{N}(0, 1/m)$ entries,
and $\x$ be a $K$-sparse $\alpha$-strongly-decaying signal. Then \eqref{e:prob} holds with
$\phi(t)$ being defined in \eqref{e:phit}.
\end{corollary}

Since $\phi(t)$ in \eqref{e:phit} is much smaller than $t$ when $t$ and/or $\alpha$
is large (see Fig. \ref{f:phi1}),
the right-hand side of \eqref{e:prob} with $\phi(t)$ being defined in \eqref{e:phit}
can be much larger than that with $\phi(t)=t$.
This essentially implies that $\mathbb{P}(\mathbb{S})$ is larger for recovering $\alpha$-strongly-decaying sparse signals
than that for recovering  flat sparse signals (i.e., the magnitudes of all the nonzero entries are identical). More details on this are given in Sec. \ref{s:sim}.

Finally, we consider the recovery of random signals $\x$.
Specifically, we assume that $\x$ is $K$-sparse with $\x_{\Omega}\sim\mathcal{N}(0, \sigma^2\I)$
for certain $\sigma$.
This kind of sparse signals arise from many applications, such as sparse activity users detection
\cite{CheSY18}.
If $\x_{\Omega}\sim\mathcal{N}(0, \sigma^2\I)$, then $\x_{\Omega}/\sigma\sim\mathcal{N}(0, \I)$.
Since $\frac{\|\x_{\mathcal{S}}\|_{1}}{\|\x_{\mathcal{S}}\|_{2}}
=\frac{\|\x_{\mathcal{S}}/\sigma\|_{1}}{\|\x_{\mathcal{S}}/\sigma\|_{2}}$,
to find a function $\phi(t)$ such that \eqref{e:csk} holds for
$K$-sparse signal $\x$ satisfying $\x_{\Omega}\sim\mathcal{N}(0, \sigma^2\I)$,
we only need to find a $\phi(t)$ such that \eqref{e:csk} holds for
$K$-sparse $\x$ satisfying $\x_{\Omega}\sim\mathcal{N}(0, \I)$.
Since $\x$ is a random signal, it is impossible to find a  $\phi(t)$ such that \eqref{e:csk}
always holds. But we can find a $\phi(t)$ such that \eqref{e:csk} holds with high probability.

If $\x_{\Omega}\sim\mathcal{N}(0, \I)$, then the expected value of $\|\x\|_{1}^2$
divided by the expected value of $\|\x\|_{2}^2$ equals to $\sqrt{2/\pi}|\Omega|$.
Therefore,  we may  try $\phi(t)=\sqrt{2/\pi}t$.
However, from simulations, we found that \eqref{e:csk}
does not hold with high probability when $|\Omega|$ is small.
Fortunately, $\phi(t)$ defined in \eqref{e:phit2} below is a suitable function to ensure \eqref{e:csk} holds with high probability:
\begin{align}
\label{e:phit2}
\phi(t)=
\begin{cases}
      0.8\,t & t\geq 30 \\
      24 & 25\leq t\leq 29\\
      t & t\leq24
   \end{cases}.
\end{align}

Fig. \ref{f:rand2} shows the probability of  $\|\x\|_{1}^2/\|\x\|_{2}^2\leq \phi(t)$
for $t=1,2\ldots,50$ over 50000 realizations, where $\phi(t)$ is defined in \eqref{e:phit2}.
From Fig. \ref{f:rand2}, one can see that \eqref{e:csk} holds with $\phi(t)$ in \eqref{e:phit2}
with probability larger than $0.996$. Hence, we have the following observation:


\begin{figure}[t]
\centering
\includegraphics[width=3.0in]{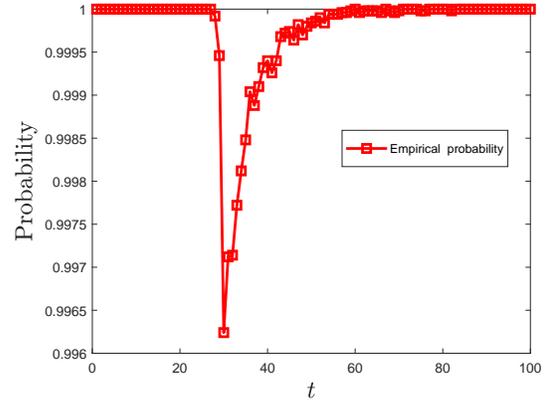}
\caption{The empirical probability of $\|\x\|_{1}^2/\|\x\|_{2}^2\leq \phi(t)$  over 50000 realizations of $\x\in \mathbb{R}^{t}\sim\mathcal{N}(0, \I)$}
\label{f:rand2}
\end{figure}

\begin{observation}
\label{c:Gauss}
Suppose that $\A\in \mathbb{R}^{m\times n}$ is a random matrix with i.i.d. $\mathcal{N}(0, 1/m)$ entries,
and $\x$ is a $K$-sparse signal satisfying $\x_{\Omega}\sim\mathcal{N}(0, \sigma^2\I)$ for
certain $\sigma$. Then \eqref{e:prob} holds with
$\phi(t)$ being defined in \eqref{e:phit2} with empirical probability larger than $0.996$.
\end{observation}

Fig. \ref{f:rand2} indicates that \eqref{e:csk} holds with
$\phi(t)$ being defined in \eqref{e:phit2} with probability larger than $0.996$.
Since $\phi(t)$ being defined in \eqref{e:phit2} is much smaller than $\phi(t)=t$ for large $t$,
Observation \ref{c:Gauss} essentially implies that $\mathbb{P}(\mathbb{S})$ is larger for recovering Gaussian
sparse signals than that for recovering  flat sparse signals.
More details on this will be provided in Sec. \ref{s:sim}.

\section{Simulation tests}
\label{s:sim}

This section performs simulations to illustrate Theorem \ref{t:noiseless},
Corollaries \ref{c:regular}--\ref{c:decay} and Observation \ref{c:Gauss} and compare them with \cite[Theorem 6]{TroG07b}.

We generated 1000 realizations of linear model \eqref{e:model}.
More specifically, for each fixed $m$, $n$ and $K$, for each realization,
we generated an $\A\in \mathbb{R}^{m\times n}$ with i.i.d. $\mathcal{N}(0, 1/m)$ entries;
we randomly selected $K$ elements from the set $\{1,2,\ldots, n\}$ to form the support $\Omega$ of $\x$;
and then generated an $\x\in \mathbb{R}^{n}$ according to the following four cases:
1) $x_i=1$ for $i\in \Omega$ and $x_i=0$ for $i\notin \Omega$;
2) The $i$-th element of $\x_{\Omega}$ is $1.1^{K-i}$ for $i\in \Omega$ and $x_i=0$
for $i\notin \Omega$;
3) The $i$-th element of $\x_{\Omega}$ is $1.2^{K-i}$ for $i\in \Omega$ and $x_i=0$
for $i\notin \Omega$;
4) $\x_{\Omega}=\mbox{randn}(K,1)$ and $\x_{\Omega^c}=\0$, where randn is a MATLAB built-in function.
After generating $\A$ and $\x$, we set $\y=\A\x$.
Then, we use OMP to reconstruct $\x$,
and denote the number of exactly recovery of $\x$ (note that $\x$ is thought as exactly recovered if
the 2-norm of the difference between the returned $\x$ and generated $\x$ is not larger than $10^{-10}$)
over 1000 as ``Empirical".

\begin{figure}[!t]
\centering
\includegraphics[width=2.9in]{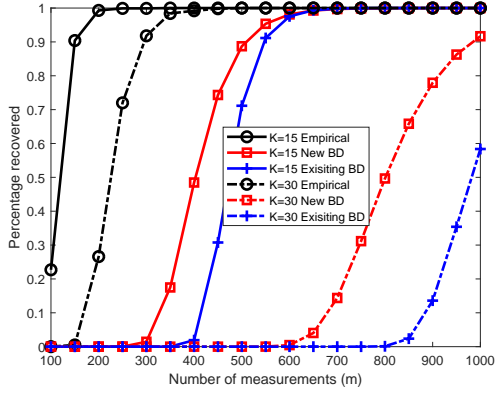}
\caption{Recovery of $K$-sparse flat signals}
\label{f:prob1}
\end{figure}
\begin{figure}[!t]
\centering
\includegraphics[width=2.9in]{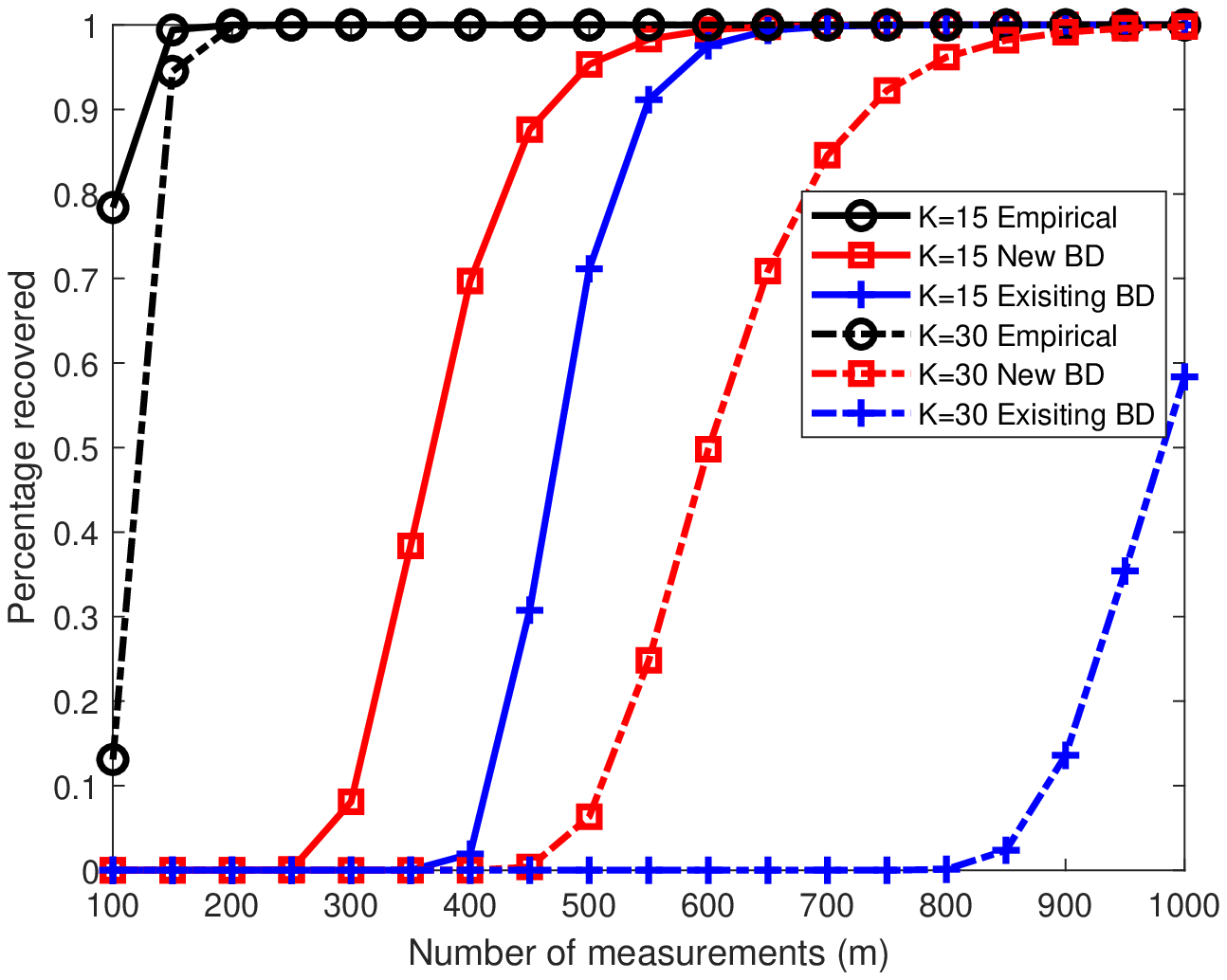}
\caption{Recovery of $K$-sparse $1.1$-strongly-decaying signals}
\label{f:prob21}
\end{figure}
\begin{figure}[!t]
\centering
\includegraphics[width=2.9in]{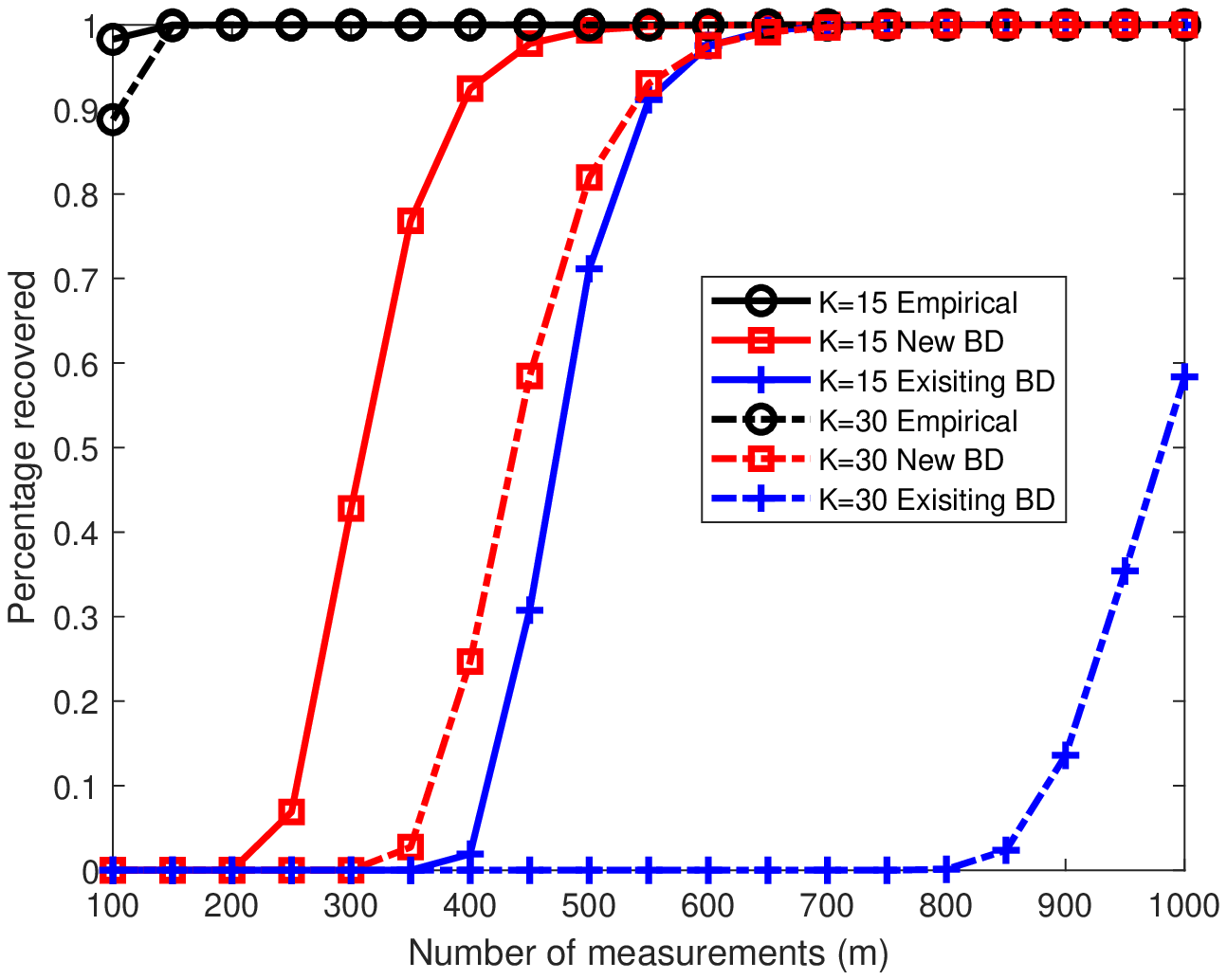}
\caption{Recovery of $K$-sparse $1.2$-strongly-decaying signals}
\label{f:prob22}
\end{figure}
\begin{figure}[!t]
\centering
\includegraphics[width=2.9in]{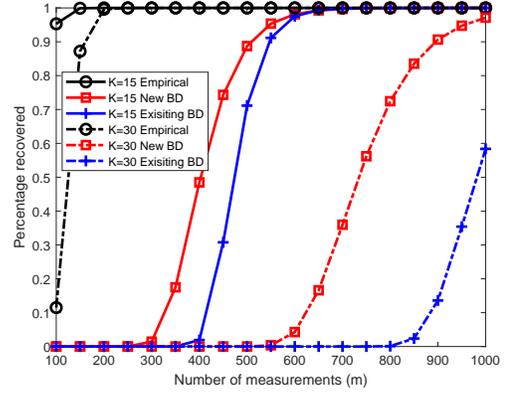}
\caption{Recovery of $K$-sparse Gaussian signals}
\label{f:prob3}
\end{figure}
We respectively compute the right-hand side of \eqref{e:prob} with $\phi(t)=t$, $\phi(t)$
defined by \eqref{e:phit}  with $\alpha=1.1$ and $\alpha=1.2$ and $\phi(t)$ defined
by \eqref{e:phit2} for the four cases, and  denote them as ``New BD".
To compare Corollaries \ref{c:regular}--\ref{c:decay} and Observation \ref{c:Gauss} with \cite[Theorem 6]{TroG07b},
we also compute the right-hand side of \eqref{e:probtrop} and denote it as ``Existing BD".
Since the lower bound on $\mathbb{P}(\mathbb{S})$ given by \cite[Theorem 6]{TroG07b} uses the sparsity of $\x$ only,
``Existing BD" are the same for all the four cases.

Figs. \ref{f:prob1}-\ref{f:prob3} respectively display ``Empirical", ``New BD" and ``Existing BD"
for $m=100:50:1000$ and $n=1024$ with $K=15$ and $K=30$ for $\x$ from cases 1-4.
Figs. \ref{f:prob1}-\ref{f:prob3} show that ``New BD" are much larger than ``Existing BD"
for all the four cases which indicates that the lower bounds on $\mathbb{P}(\mathbb{S})$ given by
Corollaries \ref{c:regular}--\ref{c:decay} and Observation \ref{c:Gauss} are much sharper than that given by \cite[Theorem 6]{TroG07b}.
They also show that OMP  has significantly better recovery performance in recovering
$\alpha$-strongly-decaying and Gaussian sparse signals than recovering flat sparse signals,
and the recovery performance of the OMP algorithm
for recovering $\alpha$-strongly-decaying sparse signals becomes better as $\alpha$ gets larger.


\vspace{-5mm}
\section{Conclusions}
\label{s:con}

In this paper, we developed lower bounds on the probability of exact recovery using $K$ iterations of OMP
for $\x$ satisfying a condition that characterizes the variations in the magnitudes of the nonzero
entries of $\x$.


\bibliographystyle{IEEEtran}
\bibliography{ref-RIP}

\end{document}